\documentclass[10pt,twocolumn]{article}
\usepackage{amsmath,amssymb}
\usepackage{graphicx}
\usepackage{latexsym}

\DeclareMathDelimiter{\lparen}{\mathopen} {operators}{"28}{largesymbols}{"00}
\DeclareMathDelimiter{\rparen}{\mathclose}{operators}{"29}{largesymbols}{"01}
\begingroup
\catcode`\(=\active \catcode`\)=\active
\gdef({\left\lparen}
\gdef){\right\rparen}
\endgroup

\begin{document}

\title{The Effect of Network-Topology to Propagation on Networks  }
\author{Norihito Toyota, Tomoharu Sakamoto and Fumiho Ogura \and Hokkaido Information University, Ebetsu, Nisinopporo 59-2, Japan \and email :toyota@do-johodai.ac.jp }

\maketitle
\abstract{%
We study the effect of the network topology to propagation phenomena on networks in this article. 
We do not assume any propagation model such as the contact process or  SIR model\cite{Ker} because the study is only 
the consideratons of the purely topological effect, especially the effect of cycles of a network. 
To uncover universal properties independent of explicit propagation models is expected due to it.  
First of all,  we introduce some indeces for propagation phenomena of a network. 
Second we introduce a concept of cycles with a little differences to usal cycles, which is called "STOC" in the body of this article. 
We find some analytic relations between thesm, STOC and some indeces.   
Moreover we can find the total number of STOCs in a network, analytically. 
This consideration leads to  an extension of the celebrated "Euler's polyhedron formula", which is only applicable to planar graphs. 
This extended formula is applicable to any graphs. 
Last we estimate numerically the indeces and the number of STOCs based on the theoretical considerations for some complex networks and make some discussion on the effects of cycles in networks  to propagation.     
 }
 
\date{Keywords;Small world network, Scale free network, topology, cycle, propagation, Eurer's  Polyhedron Theorem}
%\keywords{%
%Small world network, Scale free network, topology, cycle, propagation, Eurer's  Polyhedron Theorem
%}

%\maketitle
%-----------------------------------------------------------------------

\section{Introduction}

\hspace{5mm} In 1967, Milgram has made a great impact on  the world by advocating "Six Degrees of Separation"\cite{Milg}. 
After that, his research group makes some social experiments to establish the conjecture\cite{Milg2}, \cite{Milg3}.  
Recently Watts and his  research group  adduced evidence in support of it more intensively by experiments using e-mail\cite{Watt4}, \cite{Watt3}. 

\normalsize 
We have studied the mathematical structures of complex networks \cite{Toyota3}-\cite{Toyota7} to understand "six degrees of separation" advocated by Milgram\cite{Milg}. 
 Especially, we have developed an original formulation \cite{Toyota3}-\cite{Toyota7} based on " string formulation" developed by Aoyama\cite{Aoyama}   in order to study the effect of cycles in a network to propagation.   
 
\begin{figure}[tb]
\begin{center}
\includegraphics[scale=0.65]{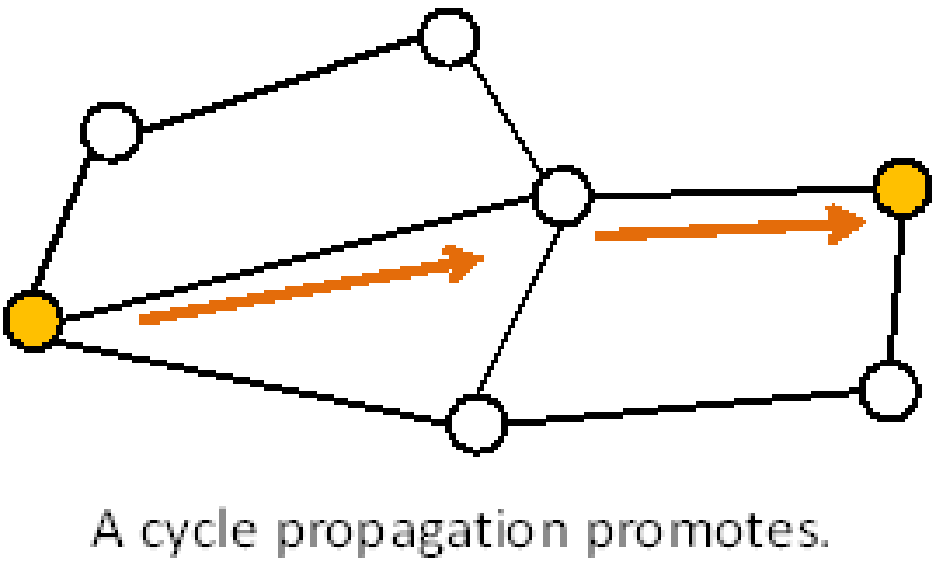}
\includegraphics[scale=0.65]{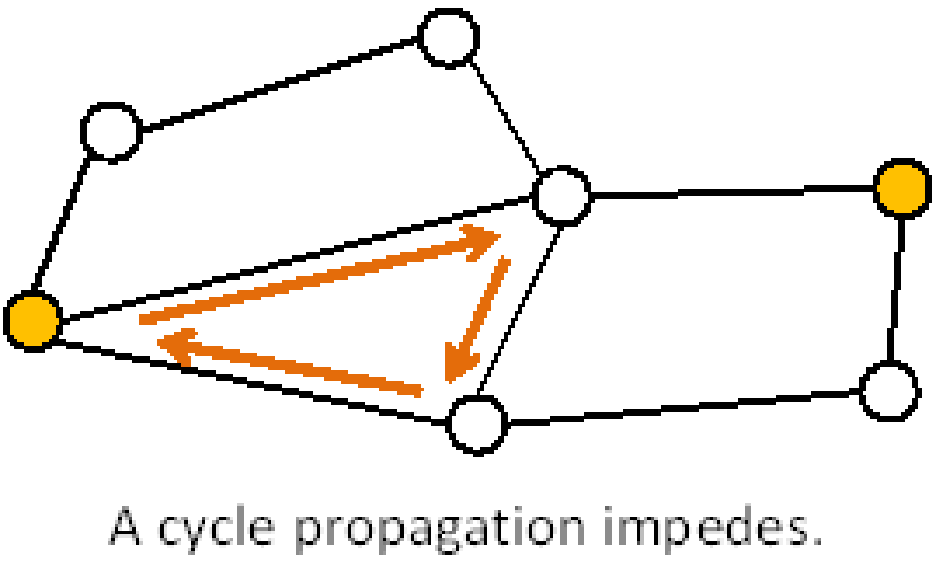}
\end{center}
\caption{Two roles of cycle structure on a graph.}
\label{50}
\end{figure}

 We have proceeded with our study for scale free networks and small world networks in the basis of Milgram condition proposed by Aoyama \cite{Aoyama}. 
As a result, it proved that the generalized clustering coefficient, which takes on the responsibility of cycles in a network,  has the opposite effect for propagation in the both networks. 
That is, the existence of closed paths in a network impedes the propagation  in small world networks, but 
does not in  scale free networks\cite{Toyota3}-\cite{Toyota7}. 
An epitom is given in Fig.\ref{50}. 
The analyses in the articles, however,  did no more than the study up to cycles with 6 nodes. 
%We cahnge the strategy for the problem to employ an approach based on an ego-network inspired from \cite{new}  in this article.
%The study of cycles with this size will be the limits due to both theoretical and numerical power of a computer. 

So changing the approach to this problem, we develop arguments from a new angle, that is  an ego-network inspired by \cite{new}, on this problem.  
  We also study propagation phenomena on networks from purely topological point of view, especially the effect of cycles of a network.  
Thus we do not assume any propagation models such as the contact process or  SIR model\cite{Ker}. 
First of all, we introduce some indeces for propagation phenomena of a network, which is easy to estimate. 
Second we introduce a concept of cycles with a little differences to usal cycles,  called "STOC".   
We find some relations between STOC and the indeces,  and one among them is represented by a recursion relation for regular networks. 
This recursion relation is solvable analytically so that the solution gives an explict relation between STOC and the indeces. 
Thanks to the recursion relation, we  also can evaluate  the total number of STOC in a network that 
leads to an extension of the celebrated "Euler's polyhedron formula", which is only applicable to planar graphs. 
Last we estimate numerically the indeces and the number of STOCs based on the theoretical considerations for some complex networks and make some discussions  on the effects of cycles in networks  to propagation.     

\section{Indeces for propagation}
\begin{figure}[tb]
\begin{center}
\includegraphics[scale=0.6]{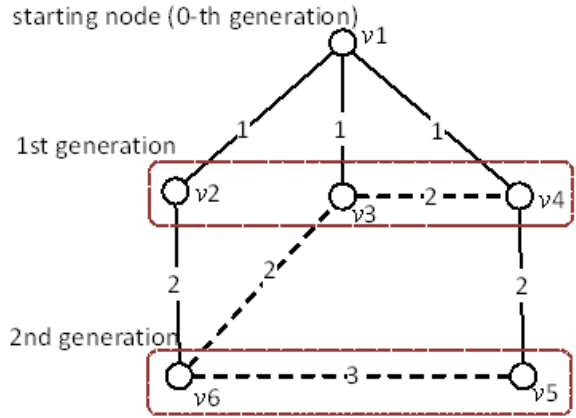}
\end{center}
\caption{The generation number of nodes and edges}
\label{100}
\end{figure}

We first introduce a generation number of nodes. 
We choose a node freely.  This node is  $0$-th generation node. 
We call the $n$-generation node that can arrive at the $0$-th generation node by shortest $n$-hops.
Thus the number of nodes of the first generation is the same as the degree $k_0$ of the $0$-th generation node. 
For example of Fig.\ref{100} where $v_1$ is the $0$-th generation node,  $v_2$ and $v_3$ are first generation nodes and 
$v_4$, $v_5$ and $v_6$ are second generation nodes, respectively. 

Though we should properly consider a rate of propagation to adjacent nodes  in introducing indeces for propagation, %\cite{masu}, 
we here introduce rather simple indeces, since our aim is to uncover the  purely topological effect to the propagation on networks. 
Indeces introduced in this article represent how many nodes the nodes at a certain generation can arrive. 
Thus the indeces in this article represent the maximum number of nodes that nodes at every generation can reach by one hop.  
It would be natural that these indeces give an estimation of propagation.  
When the starting node is $v_i$, we define the local absolute index $N _ { M } ( v _ i  )$  of $M$-generation by the number of nodes 
that newly reaches by $M$-th hop from $v_i$. 
In Fig.\ref{100}，when $v _ 1$is the starting node, we find $N _ { 0 } ( v _ 1 ) = 1$, $N _ { 1 }  (v _ 1 ) = 3$ and $ N _ { 2 }  (v _ 1 ) = 2  $, 
where the index introduced depends on the starting node.
We further introduce the absolute index $N_M$ independently of a starting node as the average over all starting nodes; 
%\begin{activeparen}
	\begin{eqnarray}
		 N _ { M } = \frac { 1 } { N } \sum _ { i = 1 } ^ { N }  N _ { M } ( v _ i ), 
		\label{e60}
	\end{eqnarray}
%\end{activeparen}
where $N$ is the total number of nodes in the considering network. 
These indeces means the maximum number of nodes that can propagate  from a generation to the next generation.  
The expected value of the absolute index also gives the average length from a starting node $v_i$ to nodes in a whole network. 
So the index gives the detail  for an average length.  

Moreover we define the local relative index $R_M(v_i)$ of $M$-generation by the ratio between the local absolute index for $M$-th generation and the one 
for $(M+1)$-th generation; 
%\begin{activeparen}
	\begin{eqnarray}
		R_ { M } ( v _ i ) = \frac{ { N } _ { M + 1 } (v_i) } { { N } _ { M } ( v _ i  ) }.
		\label{e70}
	\end{eqnarray}
%\end{activeparen}
%
The correspondong relative index $R_M$ is defined by the average over all starting nodes; 
%\begin{activeparen}
	\begin{eqnarray}
		R _ { M } = \frac { 1 } { N } \sum _ { i = 1 } ^ { N } R _ { M } ( v _ i ). 
		\label{e80}
	\end{eqnarray}
%\end{activeparen}
%
%
%
%%%%%%%%%%%%%%%%%%3%%%%%%%%%%%%%%%%%%%%%%%%%
\section{Definition of STOC}
%
%STOCとは，一般の閉路を簡略化したものである．
We define the generation number of edges. 
We define $n$-th generation edge as the set of edges that connect a $(n-1)$-th generation nodes to a $(n-1)$-th or a $n$-th  generation nodes. 
The figures on edges show the generation number of edges in Fig.\ref{100}.   

Next we define the primary edges that are edges connecting  $n$-th generation nodes and  $(n+1)$-th generation nodes,  
where anyone among the edges is the primary edge but others are secondary edges 
in the cases that plural $n$-th generation nodes connect a $(n+1)$-th generation node. 
Edges that connect among nodes in the same generation are also defined as secondary edges. 
Solid lines are primary edges and dotted lines are secondary ones in Fig.\ref{100}.

We define STOC as cycles that include only one secondary edge. 
A cycle  $v1-v4-v3-v1$ is STOC consisting of three nodes in Fig.\ref{100}, 
since the cycles includes only one secondary edge. 
Cycles $v1-v2-v6-v3-v1$ and $v1-v2-v6-v4-v3-v1$ are likewise  STOCs with four nodes and five nodes, respectively. 
The cycle $v4-v3-v6-v5-v4$, however, is not STOC, since it includes  two secondary nodes. 

STOC is a concept that removes redundancy in usual cycles in a sense.
 STOCs that newly is made  by introducing $M$-th generation edges are called $M$-generation STOCs.  
$C^{(j)}_{M} ( v_ i ) $ represents the number of $M$-generation STOCs with node $j$ when the starting node is $v _ i $.  
The numbers of $M$-th generation STOCs with any nodes are written as 
$C_M ( v _ i ) = ( C^{(3)}_M,C^{(4)}_M ( v _ i ), \cdots,C^{(j)}_M ( v _ i ) ,\cdots,C^{(2M)}_M ( v _ i )  )$,  
where $2M \geq j \geq 3$, because there are no cycles with nodes  less than 3 and 
there are only cycles with at most $2M$ nodes  up to $M$-th genetartion node. 
In Fig.\ref{100}，we find $C_0 ( v _ 1 ) = (0), \;C_1 ( v _ 1 ) = (0,0),\; C_2 ( v _ 1 ) = ( 1,1,0,0 ), \;C_3 ( v _ 1 ) = ( 0 , 0 , 1 , 0,0,0)$.  
%
%%%%%4444444444444444444444444444444444444444444
\section{Relations between the indeces and STOC}
\subsection{The absolute index and STOC}
\begin{figure}[tb]
\begin{center}
\includegraphics[scale=0.6]{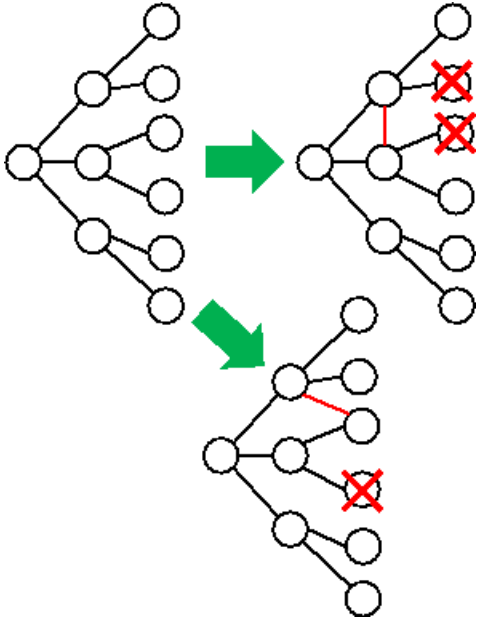}
\end{center}
\caption{Relations between cycles and propagation}
\label{150}
\end{figure}
The local absolute index in every generation becomes maximum at a tree like structure  in a network as shown in the left figure of Fig.\ref{150}. 
When there are some cycles, as shown in the  above figure to the right of Fig.\ref{150}, the number of nodes in the next generation decreases by two every cycle with odd number nodes. 
Moreover when a cycle with even number nodes is made, the number of nodes in the next generation and the further next generation decreases by one 
as is shown in the lower right network in Fig.\ref{150}, respectively.  
From these considerations, we can obtain the following relation between the absolute index $N_M(v_i)$ and STOC for $M\geq 2$. 
%
%%%%%%%%%%%%%%%%%%%%%%%%%%%%%%%%%%%%%%%%%%%%%%%%%%%%%%%%%%%%%%%%%%%%%%%%%%%%%%%%%%%%%%%%%%%%\mathindent=0zw
%\begin{activeparen}
	\begin{eqnarray}
		N_{M} ( v_i ) &= &\sum_{j=1}^{N_M-1} \bigl( k^{j}_{M-1} ( v_i ) -1\bigr) - 2 \sum_{i=3,5 \cdots}^{2M-1} C^{(i)}_{M} ( v_i ) \nonumber\\
		 & - &\sum_{i=4,6 \cdots}^{2M} C^{(i)}_{M} ( v_i ) -\sum_{i=4,6 \cdots}^{2(M-1)} C^{(i)}_{M-1} ( v_i ),
		\label{e120}
	\end{eqnarray}
%\end{activeparen}
where $k^{j}_{M-1} ( v_i )$ is the degree of $j$-th node in $M-1$ generation. 
 It can be proven that  (\ref{120} does not depend on which edges is secondary edges. 
 The proof is  omitted in this article because of limited page space. 
The essential fact for the proof is that one secondary corresponds to just one cycle.    
The first term of the right hand side in (\ref{e120}) represents the contribution to  $N_{M}$ at tree approximation of a network.  
The second term and the third term of right hand side represent the decreasing numbers of nodes in the present generation when 
cycles with odd number of nodes and even number of nodes are made, respectively.      
The fourth term represents the decreasing numbers of nodes in the present generation when 
a cycle with the even number of nodes is made at two generations before.      
(\ref{e120}) is actually confirmed with some regural networks, a ring-type network, extended rings, the triangular lattice, tetragonal lattice 
and the tetragonal lattice on torus.   

When the degree of all nodes included in the network is equal to constant $k$, we obtain
%\begin{activeparen}
	\begin{eqnarray}
		N_{M} ( v_i ) &=& N^{(k)}_{M-1} ( v_i ) (k-1) - 2 \sum_{i=3,5\cdots}^{2M-1} C^{(i)}_{M} ( v_i )\nonumber\\
		& -& \sum_{i=4,6 \cdots}^{2M} C^{(i)}_{M} ( v_i ) -\sum_{i=4,6 \cdots}^{2(M-1)} C^{(i)}_{M-1} ( v_i ). 
		\label{e121}
	\end{eqnarray}
%\end{activeparen}
%
%where $k_M$ is the average degree over the nodes of the $M$-th generation in (\ref{e120}). 

(\ref{e120}) is a recursion relation for the local absolute index. 
%また，(\ref{e120})について三角格子，正方格子，正方格子トーラス，円環，拡張版円環などのレギュラーネットワークでその正当性が確認されている．
%When every node has the same degree $k_M=k$, 
We can analytically solve the recursion relation; 
\begin{eqnarray}
	&N _ { M }  ( v _ i ) &= k(k-1)^{M-1} \nonumber \\
	&-& \sum_{j=2}^{M} \Bigl( 2 \sum_{ i = j } ^ { M } ( k - 1 ) ^ { M - i } C^{(2j-1)}_{i} ( v _ i )  \nonumber\\
	&+k&  \sum _ {  i = j } ^ { M - 1 } ( \! k \! - \! 1 \! ) ^ { M \! - \! 1 \! - \! i } C ^ { ( 2 j ) } _ { i } ( v _ i ) + C ^ { ( 2 j ) } _ { M }  ( v _ i ) \Bigr) .
	\label{e150}
\end{eqnarray}
 %%%%%%%%%%%%%%%%%%%%%%%%%%%%%%%%%%%%%%%%%%%%%%%%%%%%%%%%%%%%%%%\mathindent=1zw
%
%
(\ref{e150})  can be proven by mathematical induction and also applies for $M=1$. 
Thus we could indicate the relation between the local absolute index  $N^{(k)}_{M}$ and the number of STOCs by (\ref{e150}).  
It is considered that STOC  impedes propagation on the whole from (\ref{e150}) in networks with  the same degree for every node.
STOCs with the odd number of nodes and STOCs with the even number of ones have a different type effect on the impediment of the propagation.% from \ref{e150}.  

%We here consider networks with the constant average degree $k_M$ in every generation. 
%In general it is difficult to find the total number of cycles in the whole network  in the usual sense.
%We can, however, find the total number $S_L$ of STOCs in the whole network. 
%Evaluationg the sum of the number of STOCs from $0$0th generation upto the final generation $L$, 
%which means that all nodes in the network from the starting node $v_i$ are included up to $L$-generation , we obtain the following equation;  
%\begin{activeparen}
	%\begin{eqnarray}
%		\sum _ { i = 0 } ^ { L } N _ { i } ( v _ l ) &=& 2 + ( k_M - 1 ) \sum _ { i = 0 } ^ { L - 1 } N _ { i } ( v _ l ) \nonumber\\
%		&-& \! 2 \sum _ { i = 2 } ^ { L }  \sum _ { j = 3 } ^ { 2 i }  C ^ { ( j ) } _ { i } \! ( v _ l ) + \! \sum _ { i = 4 , 6 , 8 \cdots } ^ { 2 L } \! C ^ { ( i ) } _ { L } \! ( v _ l ) .
%		\label{e200}
%	\end{eqnarray}
%\end{activeparen}
%
\subsection{Total number of STOCs and Euler's polyhedron theorem}
It is also possible to find the total  number of STOCs in the whole network from (4). 
Let $L$ be the last generation for nodes in a network, so there are no nodes in $L+1$ generation.  
Taking the sum from $0$-th generation to $L$-th generation for (\ref{e120}),
we obatin 
 %\begin{activeparen}
	\begin{eqnarray}
		\sum_{i=0}^{L} N_{i} &=& \sum_{i=0}^{L} N_{i} + N_{L} = 2 + \sum_{i=0}^{L-1} (\overline{k_{i_{j}}}-1)N_{i} \nonumber\\
		& & - 2 \sum_{i=2}^{L} \sum_{j=3}^{2i} C^{(j)}_{i} + \sum_{i=4,6,8 \cdots}^{2L} C^{(i)}_{L+1},
		\label{e250}
	\end{eqnarray}
The third term in the right hand side of (\ref{e250}) shows the total number of STOCs. 
We must manage the last term in the right hand side of (\ref{e250}) in order to obtain the total number of STOCs 
without any other information about STOCs.  

 When there are $L$-th generation nodes, there are, however, $L+1$ generation edges. 
We formally consider $(L+1)$-th generation as a dummy in order to include $L+1$ generation edege. 
As is shown later, this device solve  the problem for the last term. 
Thus taking the sum from $0$-th generation to $(L+1)$-th generation,
we obatin 
 %\begin{activeparen}
	\begin{eqnarray}
		\sum_{i=0}^{L+1} N_{i} &=& \sum_{i=0}^{L} N_{i} + N_{L+1} = 2 + \sum_{i=0}^{L}(\overline{k_{i_{j}}}-1) N_{i} \nonumber\\
		& & - 2 \sum_{i=2}^{L+1} \sum_{j=3}^{2i} C^{(j)}_{i} + \sum_{i=4,6,8 \cdots}^{2(L+1)} C^{(i)}_{L+1},
		\label{e250}
	\end{eqnarray}
%\end{activeparen}
where $\overline{k_{i_{j}}}$ is the number of edges between the nodes in $(i-1)$-th generation and $j$-th node in $i$-th generation.
%%%%%%%%%%%%%%%%%%%%%%%%%%%%%%%%%%%%%%%%%%%%%%%%%%%%%%%%%%%%%%%%%% as shown in Fig.4.%%%FIGFIG4%%%%%%%%%%% 
The second term in the right hand side in (\ref{e250}) is just the total number of STOCs in a whole network. 
The last term of the right hand side in (\ref{e250})  vanishes, since there is no STOC with odd numbers of nodes in the 
$L+1$-th generation. 
%We here introducce  symbols  $(v_{L_{1}},v_{L_{2}}, \cdots ,v_{L_{N_{L}}})$ for nodes of $L$-th generation and 
%$(k_{L_{1}},k_{L_{2}}, \cdots ,k_{L_{N_{L}}})$ for degrees of nodes of $L$-th generation.  
%この表記を用いて$L$世代目の次数を$k$にするようにした$L+1$世代目の頂点をツリーネットワークになるように作成すると$L+1$世代目の頂点数は%以下の式のようになる．
Thus we obtain the total number of STOCs 
%\begin{activeparen}
	%\begin{eqnarray}
%		N_{L+1} = \sum_{i=1}^{N_{L}}(k_M-k_{L_{i}}) = k_M N_{L} - \sum_{i=1}^{N_{L}} k_{L_{i}}.
	%	\label{e230}
	%\end{eqnarray}
%\end{activeparen}
%
%Substituting (\ref{e230}) into $N_{L+1}$ in (\ref{e250})，we obtain 
%
%\begin{activeparen}
	\begin{eqnarray}
			S _ L ( v _ i) &=& \sum _ { i =  2 } ^ { L } \sum _ { j = 3 } ^ { 2 i }  C ^ { ( j ) } _ { i } \! ( v _i ) \! \nonumber \\
			&=& 1+\frac { 1 } { 2 }   \sum _ { i = 0 } ^ { L } N _ { i }  \overline{k_{i_{j}}}  - \sum_{i=0}^{M} N _ { i} .  % \nonumber\\
%			&=& \frac { 1 } { 2 } ( \! 2 \! + \! ( \! k _ M \! - \! 2 \! ) N  \! - \!  k _ M N _ { L }   \! + \!  \sum _ { i = 1 } ^ { N _ { L } } k _ { L _ { i } } \! ).
			\label{e300}
	\end{eqnarray}
%\end{activeparen}
%
The fact is, this formula is applicable not only to the last generation $L$, but also to every generation $M$ by simirarly thinking some dummies   
at every generation. 
Thus we can estimate the total number $S _M ( v _ i )$ of STOCs upto any generations, where $\overline{k_{i_{j}}}$ is numerically calculated in actuality.  

Further we find that the (\ref{e300}) leads to an exatension  of the celebrated Euler's polyhedron theorem (without genuses) at the last generation, 
because the second term and the third term of the right hand side in   (\ref{e300}) means the total number of edges and nodes in a whole network, respectively; 
	\begin{eqnarray}
	Euler:  surface &=& 1 +  edges - nodes,	\label{e330}\\
	STOC:S _ M(v_i) &=& %1 + \frac{ 1 }{ 2 } \sum_{i=0}^{M} \sum_{j=1}^{ N _ { i } } \overline{ k _ { i _ { j } } } - \sum_{i=0}^{M} N_{i}
	1 +   edges - nodes, \label{e350}
	\end{eqnarray}
where   (\ref{e330}) applies when one surface on the outside of a network is omitted and $S _ M(v_i)  $ in (\ref{e350}) does not depend on any starting node $v_i$, 
because it is determined only by the total numbers of the surfaces and the edges in the considering network.    
While the Euler's polyhedron theorem  applies only to planar graphs,  (\ref{e350}) applies in any graphs. 
Thus when we put up a menbrane at one STOC, (\ref{e350}) gives an exatension  of the Euler's polyhedron theorem. 
%aaaaaaaaaaaaaaaaaaaaaaaaaaaaaaaaaaaaaaaaaaaaaaaaaaaaa
%%%%%%%%%SSS55555555555555555555555%%%%%%%%%%%%
\section{Simulation and its results}
We numerically calculate the number of STOCs and the absolute index in every generation in two types of complex networks to consider their relations.
One is the small world network proposed by Watts-Strogatz\cite{ws} where the probability $p$ replacing edges is a control parameter for  
the number of small size cycles.
Second is the scale free network proposed by Holme-Kim\cite{hk} where $q$, the probability making cycles consist of  three nodes, plays role of a control parameter  for the number of small size cycles. 
The network becomes Barabasi-Albert model\cite{ba} with scaling exponent $3$ at  $q=0$. 
We set up the network size $N=3000$ and the average deggree $<k>=6$ in both networks.    
The number of STOCs in every generation  is estimated by taking  a difference of two adjacent generations in (\ref{e300}) 
and the average over all starting nodes $ v _ i $.  
Ten times of average are taken at every simulation.  
The total number of STOCs  has not any effect on propagation phenomena,  since it is determined by the total number of nodes and edges in a network 
by the extended Euler's polyhedron theorem.   
We will show only the results of the absolute indeces in this article, since the relative indces are essentially the same behaviors as the absolute ones in both networks.

%但し，$ k _ M $ の代わりにネットワーク全体の平均次数を用いた．

%
\subsection{Small World Network}
\begin{figure}[tb]
\begin{center}
\includegraphics[scale=0.85]{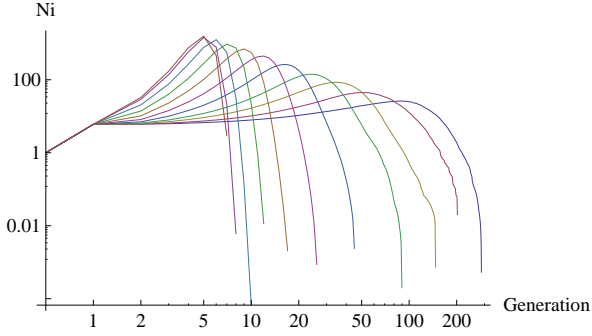}
\end{center}
\caption{The absolute index of small world network}
\label{170}
\end{figure}
\begin{figure}[tb]
\begin{center}
\includegraphics[scale=0.85]{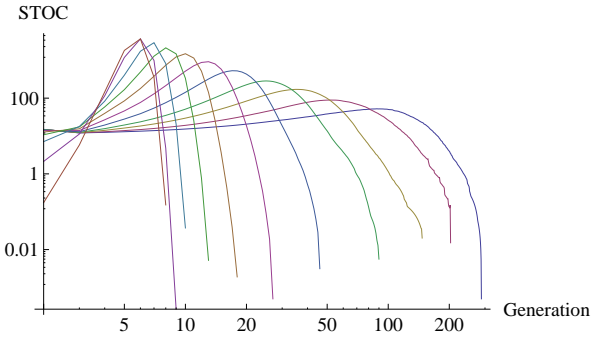}
\end{center}
\caption{STOC of small world network}
\label{200}
\end{figure}
%
%\begin{figure}[tb]
%	\begin{center}
%		\makeatletter
%		\def\@captype{table}
%		\makeatother
%		\caption{スモールワールドネットワークのSTOC}
%		\begin{tabular}{|c|c|} \hline
%			$p$ & STOC \\ \hline
%			0.1 & 6002.533 \\ \hline
%			0.2 & 6010.4413 \\ \hline
%			0.3 & 6017.6251 \\ \hline
%			0.4 & 6029.516 \\ \hline
%			0.5 & 6020.015317 \\ \hline
%			0.6 & 6030.816933 \\ \hline
%			0.7 & 6047.76075 \\ \hline
%			0.8 & 6059.89175 \\ \hline
%			0.9 & 6063.236817 \\ \hline
%			1 & 6062.486417 \\ \hline
%		\end{tabular}
%		\label{h100}
%	\end{center}
%\end{figure}
%
In the small world network, $p$ is changed from 0 to 1 by $p=2^{-m}$ for $m=0,1,2,\cdots 10$. 
The absolute index and the total number of STOCs of small world network are given by Fig.\ref{170} and Fig.\ref{200}. 
Each curved line in Fig.\ref{170} and Fig.\ref{200} shows the results of 11 kinds of $p$, where the curved line with the smallest maximum is 
one for the smallest $p$ and the one with the largest maximum is the one for $p=1.0$ in the both figures.       
These show that  Fig.\ref{170} behaves the same way as Fig.\ref{200}. 
While  from Fig.\ref{170}, one can reach many nodes in small generation at large $p$, one can reach many nodes in large generation at small $p$. 
%These results also means that STOC is an important concept for studying propagation phenomena on networks.  
%%%%%%%%%%%%hkhkhkhkh%hkhkhkhkh%hkhkhkhkh
\subsection{Holme-Kim Network}
\begin{figure}[tb]
\begin{center}
\includegraphics[scale=0.85]{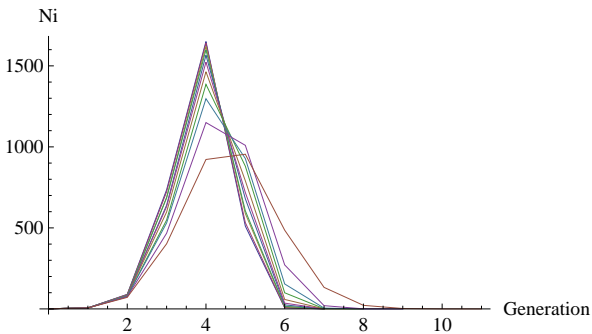}
\end{center}
\caption{The absolute index of Holme-Kim model}
\label{250}
\end{figure}
\begin{figure}[tb]
\begin{center}
\includegraphics[scale=0.85]{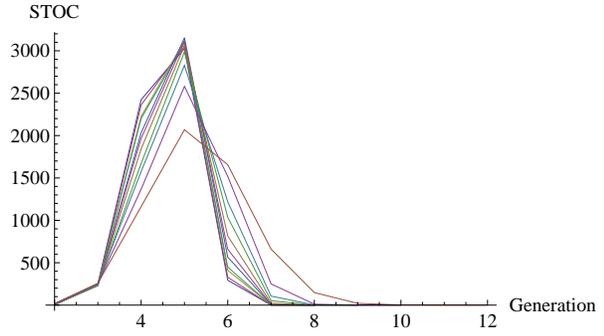}
\end{center}
\caption{STOC of Holme-Kim model}
\label{300}
\end{figure}
%
%\begin{figure}[tb]
%\begin{center}
%\includegraphics[scale=0.45]{HK_pk.eps}
%\end{center}
%\caption{Holme-Kimモデルの次数分布}
%\label{400}
%\end{figure}
%

$q$ is moved from 0 to 1  at 0.1 intervals. 
The absolute index and the total number of STOCs of Holme-Kim network are given by Fig.\ref{250} and Fig.\ref{300}. 
Each curved line in Fig.\ref{250} and Fig.\ref{300} shows the results of 11 kinds of $q$,   
where the curved line with the smallest maximum is one for the smallest $q=0$ and the one with the largest maximum is the one for $q=1.0$.      
These show that  Fig.\ref{250} behaves the same way as Fig.\ref{300}. 
While  from Fig. \ref{250}, one can reach many nodes in small generation at large $q$, one can reach many nodes in relatively large generation at small $q$. 

\subsection{Small World Network vs. Holme-Kim network}
Comparing the results in the  both networks, the maxmum of $N_i$ in Holme-Kim model is larger than the one of Small-world networks. 
We find that there is a generation reachable to many nodes of the next generation in Holme-Kim model.  
The fact is consistent with the well-known fact that Holme-Kim model is more smallnes than   Watts-Strogatz mode\cite{new}. 

From these four figures, we find that the number of STOCs reach the peak, after the absolute indeces reach the peak  in bothy networks. 
We guess that the reason is because edges  between the same generation  are also likely drawn after many nodes are connected  in a generation.   
We can  check that the total numbers of STOCS are consistent with the extended Euler's polyhedron theorem. 
From these numerical analyses, we resulted in the conclusion that STOC is an important concept for studying propagation phenomena on networks.  
%
%\begin{figure}[htbp]
%	\begin{center}
%		\makeatletter
%		\def\@captype{table}
%		\makeatother
%		\caption{Holme-KimモデルのSTOC}
%		\begin{tabular}{|c|c|} \hline
%			$q$ & STOC \\ \hline
%			0 & 6145.597639 \\ \hline
%			0.1 & 6124.972088 \\ \hline
%			0.2 & 6102.789803 \\ \hline
%			0.3 & 6091.089885 \\ \hline
%			0.4 & 6084.008633 \\ \hline
%			0.5 & 6074.826043 \\ \hline
%			0.6 & 6066.679356 \\ \hline
%			0.7 & 6045.422274 \\ \hline
%			0.8 & 6073.708108 \\ \hline
%			0.9 & 6008.338529 \\ \hline
%			1 & 6030.345998 \\ \hline
%		\end{tabular}
%		\label{h200}
%	\end{center}
%\end{figure}
%
%%%%%6666666666666666666666666
\section{Concluding remarks}
In this research, we can succeed in  making a connection with cycles with any sizes,  where cycles with only six size have been discussed  till now, 
and propagation on a network by introducing new indeces for propagation and STOCs that is a new concept of the cycles.   

%We indicate some relations between the  index for a guide of propagation and STOC, which represent a typical cycle, explictly. 
 Moreover we showed that the total number of STOCs leads to an exatension  of the celebrated Euler's polyhedron theorem 
 where surfaces are considered as menbranes put up at every STOC. 
 Thus the extended theorm applies to non-planar graph. 
 The constant $1$ appeared in (\ref{e330} is essentially a topological constant.
It is  important problem that the constant $1$ appeared in (\ref{e350} has aany topological meanings.
 
We calculated indeces and the number of STOC in every generation in some typical complex networks. 
As result, we find that the behaviors of the indeces for propagation are totally different in scale free networks and Watts-Strogatz\cite{ws}-type small world network.  
So a series of the absolute indeces  $N_M$ or the relative indeces $R_M$ may give an index characterizing various type networks. 
For it,  it is needed to study more diverse networks.  
The total number of STOCs has not any effect on propagation in networks, because it is given by the extended Euler's theorem, 
independently of details of network topology.  
We, however, resulted in the conclusion that STOC in every generation plays an important role of  propagation phenomena on networks.  
%

%%%%%%%%%%%%%%%%% BIBLIOGRAPHY IN THE LaTeX file !!!!! %%%%%%%%%%%%%%%%%%%%%%

\end{document}